\documentclass[12pt,onecolumn]{article}
\usepackage[latin1]{inputenc}
\usepackage{amsmath,amsfonts,amssymb,graphicx}
\usepackage{authblk}
\usepackage{placeins}
\usepackage{cite}

\begin{document}

\title{
Stochastic resonance in three-neuron motifs
}

\author[1]{Patrick Krauss}
\author[1]{Karin Prebeck}
\author[1]{Achim Schilling}
\author[1,2]{Claus Metzner}
\affil[1]{Experimental Otolaryngology, Neuroscience Group, University Hospital Erlangen, Friedrich-Alexander University Erlangen-N\"urnberg (FAU), Germany}
\affil[2]{Department of Physics, Biophysics Group, Friedrich-Alexander University Erlangen-N\"urnberg (FAU), Germany}

\maketitle

Corresponding author: \\
Dr. Patrick Krauss  \\
Experimental Otolaryngology \\
Neuroscience Group \\
Friedrich-Alexander University of Erlangen-N\"urnberg \\
Waldstrasse 1 \\
91054 Erlangen, Germany \\
Phone:  +49 9131 8543853 \\
E-Mail: patrick.krauss@uk-erlangen.de \\ \\


\newpage


\begin{abstract}
Stochastic resonance is a non-linear phenomenon, in which the sensitivity of signal detectors can be enhanced by adding random noise to the detector input. Here, we demonstrate that noise can also improve the information flux in recurrent neural networks. In particular, we show for the case of three-neuron motifs that the mutual information between successive network states can be maximized by adding a suitable amount of noise to the neuron inputs. This striking result suggests that noise in the brain may not be a problem that  needs to be suppressed, but indeed a resource that is dynamically regulated in order to optimize information processing.
\end{abstract}

\newpage

\section*{Introduction}

Recurrent neural networks (RNN) with apparently random connection weights are not only ubiquitous in the brain \cite{middleton2000basal, song2005highly}, but recently gain popularity in bio-inspired approaches of neural information processing, such as reservoir computing \cite{lukovsevivcius2009reservoir, verstraeten2007experimental, schrauwen2007overview}. Due to the built-in feed-back loop that distinguishes RNNs from networks with a pure feed-forward structure, input information entering a RNN at some point in time can 'circulate' within the network for extended periods. Moreover, a typical RNN will not simply conserve the input information in its original form, but transform it to new and possibly more useful representations in each iteration of the loop. This ability of RNNs to dynamically store and continuously re-code information, as well as the possibility to combine the circulating information with new inputs, is essential for the processing of sequential data \cite{skowronski2007automatic}.

From an engineering point of view, a system that stores information by continuous re-coding has to meet two requirements: (1) The number of different codes that can be represented in the system should be as large as possible, in order to enhance the probability that some of these codes are actually useful for the further information processing. (2) The transformations from one code to the next should be as reproducible as possible, since otherwise information will be lost. Taken together, these two requirements are equivalent to maximizing the mutual information (MI) between subsequent states of the system.

While the transitions between subsequent network states can be made perfectly reproducible in an artificial network of deterministic neurons, it is unclear how this reproducibility is achieved in the brain, where neural computations are subject to a large degree of internal and external noise \cite{faisal2008noise, rolls2010noisy}. 

We therefore investigate in this work how the MI between subsequent network states depends on the level of noise added to the inputs of each neuron. For this purpose, we use one of the simplest examples of RNNs, namely the class of probabilistic three-neuron motifs with ternary connection strength (-1,0,+1) \cite{krauss2018analysis}. 

Strikingly, we find that the MI is not in general decreasing monotonically with the noise level, but has for certain types of motifs a peak at some optimal level of added noise. This behaviour strongly resembles the phenomenon of stochastic resonance \cite{benzi1981mechanism, wiesenfeld1995stochastic, gammaitoni1998stochastic, moss2004stochastic}, in which adding noise to the input of a sensor can enable this sensor to recognize weak signals that would otherwise remain below the detection threshold \cite{krauss2017adaptive}. In previous studies we argued that stochastic resonance might be a major processing principle of the auditory system \cite{krauss2016stochastic} and the cortex \cite{krauss2018cross}. Based on the here presented new results, we speculate that the brain may use noise actively and in a controlled way in order to optimize information processing in its recurrent neural networks.

\newpage

\section*{Methods}

\subsection*{Three-neuron motifs}
Our study is based on Boltzmann neurons \cite{hinton1983optimal} without bias. The {\bf total input} $z_i(t)$ of neuron $i$ is calculated as:
\begin{equation}
	\mathrm{z_i(\textit{t})} = \sum\limits_{j=1}^{N} w_{ij}\;s_j(t)\,
\label{totIn}
\end{equation} 
where $s_j(t) \in\{0,1\}$ is the binary state of neuron $j$ at present time $t$, and $w_{ij}$ is the {\bf connection weight} from neuron $j$ to neuron $i$.
Depending on the total input, the {\bf on-state probability} of neuron $i$ at the next time step $t\!+\!1$ is then given by: 
\begin{equation}
	p\left( \;s_i(t\!+\!1)\!=\!1 \;\right) = \frac{1}{1\;+\;e^{-z_i(t)}}.
\end{equation}

We restrict our investigation to  network motifs that can be built from three Boltzmann neurons of the above type, using discrete, ternary connection weights $w_{ij}\in\{-1,0,+1\}$, where self connections $w_{ii}$ are permitted. 

In the following, we denote the {\bf present global state} of a three-node motif (time $t$) by the binary vector
\begin{equation}
    \vec{x}=
    \left(
    s_1(t),s_2(t),s_3(t)
    \right)
\end{equation}
and its {\bf next global state} (time step $t\!+\!1$) by
\begin{equation}
    \vec{y}=
    \left(
    s_1(t\!+\!1),s_2(t\!+\!1),s_3(t\!+\!1)
    \right)
\end{equation}
The update $\vec{x}\rightarrow\vec{y}$ is performed simultaneously for all neurons.

\subsection*{Statistical properties}

From the $3\times 3$ weight matrix $W=(w_{ij})$ of a given three-neuron motif, we first compute the pairwise {\bf transition probabilities} $p(\vec{y} | \vec{x})$ from state $\vec{x}$ to state $\vec{y}$. The resulting $8\times 8$ matrix can be interpreted as the transition matrix of a Markov process. We also determine the {\bf state probabilities} $p(\vec{x})$ for all 8 global network states in the stationary equilibrium situation.
Finally, we compute the {\bf state pair probabilities}
\begin{equation}
  p(\vec{x},\vec{y}) = p(\vec{y} | \vec{x}) \; p(\vec{x}),
\end{equation}
which completely describe the statistical (dynamical) properties of the three-node motif. 

In the case of undisturbed motifs, all statistical properties can be computed analytically. With added noise, however, we have determined these quantities numerically by analyzing long simulated time series of motif states.

\subsection*{Information theoretic properties}

Based on the above statistical properties of a given motif, we obtain the {\bf state entropy}
\begin{equation}
H(X) = -\sum_{\vec{x}} p(\vec{x}) \log_2 p(\vec{x}),
\end{equation}
where $\vec{x}$ runs through the eight possible states $(0,\!0,\!0)$, $(0,\!0,\!1)$, $(0,\!1,\!0)$, $(0,\!1,\!1)$, $\ldots \; (1,\!1,\!1)$. The state entropy measures the average amount of information circulating in the motif and is ranging from 0 to 3 bit in our case. Since this quantity does not depend on the time step, we have $H(X)=H(Y)$.

Furthermore, we obtain the {\bf mutual information} (MI) between successive states
\begin{equation}
I(X;Y) = \sum_{\vec{x}} \sum_{\vec{y}} \; p(\vec{x},\vec{y}) \log_2 \left( 
\frac{p(\vec{x},\vec{y})}{p(\vec{x})\; p(\vec{y})}
\right),
\end{equation}
which can be interpreted as the '{\bf information flux}' within the network. In our case, the MI ranges between 0 and $H(X)\!=\!H(Y)$. 

Maximizing the MI requires (1) that the amount of information circulating in the recurrent network is large, and (2) that information is transmitted with only small loss from one time step to the next. This becomes apparent in the equation 
\begin{eqnarray}
I(X;Y) &=& H(Y) - H(Y|X)\nonumber\\ 
&=& H(X) -  H(Y|X),
\label{MI_contributions}
\end{eqnarray}
where the '{\bf stochastic information loss}' $H(Y|X)$ measures the variety of output states that can follow after a given input state. In a purely deterministic system, one would have $H(Y|X)=0$. However, due to the probabilistic Boltzmann neurons and the resulting stochastic nature of the state-to-state transitions, we expect $H(Y|X)>0$.
Equation (\ref{MI_contributions}) thus states that the MI is the state entropy minus the stochastic information loss. Both quantities on the right side of this equation can be affected, to different degrees, by the presence of noise.

\subsection*{Added noise}

To simulate the effect of external noise on the information flux, continuous random values $\eta_j(t)$ are added to the internal states of the neurons $j\!=\!1\!\ldots\!3$ in every time step:
\begin{equation}
    z_j(t) \rightarrow z_j(t) + \eta_j(t).
\end{equation}
All $\eta_j(t)$ are drawn, independently, from a Gaussian distribution with zero mean and a prescribed standard deviation $\sigma$, here called the {\bf noise level}. The randomly distorted input states then enter Eq.(\ref{totIn}) as before.

\newpage

\section*{Results}

\begin{figure}[h!]
	\centering
	\includegraphics[width=1.0\linewidth]{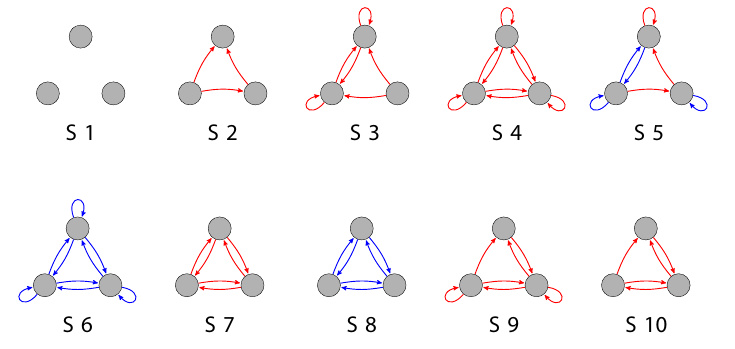}
	\caption{Examples of three-neuron motifs with ternary connections. Grey circles depict neurons, red arrows are excitatory connections ($w_{ij}=+1$), and blue arrows are inhibitory connections ($w_{ij}=-1$). In this work we consider only the motifs S1 to S4.}
	\label{motifs}
\end{figure}

There are 3411 topologically distinct three-neuron motifs with ternary connection strengths and possible self-connections \cite{krauss2018analysis}. We have investigated these motifs exhaustively with respect to their response to noise and singled out ten motifs with a particularly interesting behavior (Fig. \ref{motifs}). 

\begin{figure}[h!]
	\centering
	\includegraphics[width=1.0\linewidth]{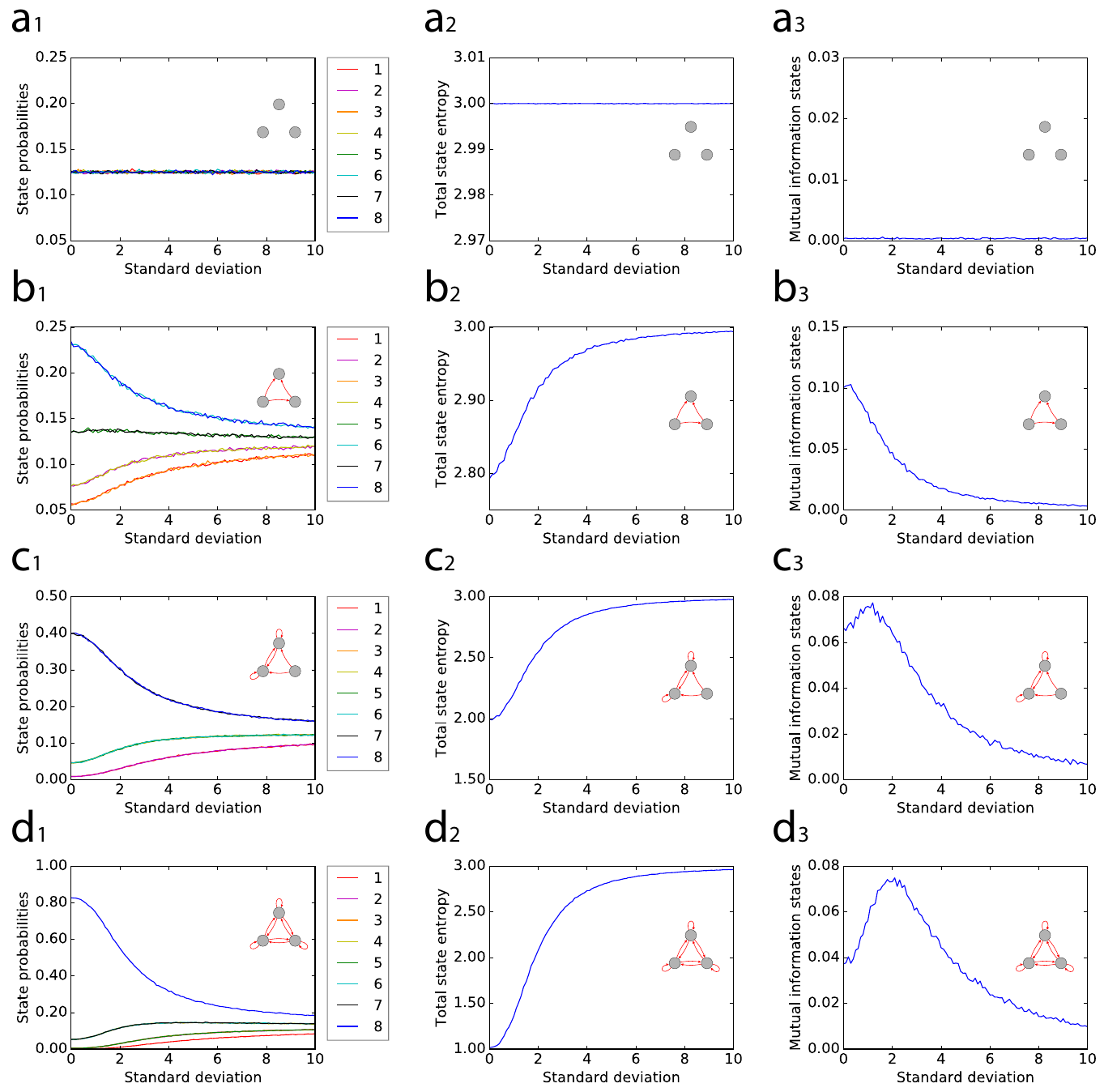}
	\caption{Statistical and information theoretical properties of selected motifs S1 to S4, as functions of the noise level $\sigma$. State probabilities $p(\vec{x})$ are shown in the left column, with the eight states labeled $1\ldots 8$ in the order of the binary number system. State entropy $H(X)$ is shown in the middle column. Mutual information (MI) of successive states $I(X;Y)$ is shown in the right column. All three quantities are averaged over $10^6$ time steps for each motif and each noise level.  
}
	\label{SR-curves}
\end{figure}

In this work, we consider only four representative motifs among these ten, denoted S1 to S4. For each of these motifs (rows of Fig. \ref{SR-curves}), we investigate the statistical and information theoretical properties as functions of the noise level $\sigma$. In particular, we compute the state probabilities $p(\vec{x})$ (left column of Fig. \ref{SR-curves}), with the eight states labeled $1\ldots 8$ in the order of the binary number system. Furthermore, we compute the state entropy $H(X)$ (middle column), and the mutual information (MI) of successive states $I(X;Y)$ (right column). All three quantities are averaged over $10^6$ time steps for each motif and each noise level. 

\paragraph*{Motif S1}

For the totally unconnected motif S1 (First row of Fig. \ref{SR-curves}), we obtain the expected results: all global states are equally probable with $p(\vec{x})\!=\!\frac{1}{8} \;\forall \vec{x}$, and the state entropy is maximal with $H(X)= 3 \;bit$. However, successive states in this motif are statistically independent and thus have no information in common, so that $I(X;Y)=0$. Furthermore, all quantities are also independent from the noise level $\sigma$, since the system is completely random from the beginning.

\paragraph*{Motif S2}

For the weakly connected motif S2 (second row of Fig. \ref{SR-curves}) and without added noise, we find that certain network states, in particular states 6 and 8, are more probable than others, leading to a sub-optimal state entropy of $H(X) \approx 2.8 \;bit$. Due to the presence of connections between the neurons, the next motif state is now to some extent predictable from the former one, so that $I(X;Y)\approx 0.1 \;bit$. As the noise level $\sigma$ is increased, all state probabilities start to asymptotically approach the uniform value of $\frac{1}{8}$. Consequently, we find a monotonous increase of the state entropy towards the maximum value of 3 bit, which in principle should favor the information flux in the motif and thus should help to increase the MI between successive states. However, the added noise also increases the stochastic information loss $H(Y|X)$ (not shown in the figure), which has a detrimental effect on the information flux. In this case of motif S2, $H(Y|X)$ is growing faster with the noise level than $H(X)$. In the end effect, this leads to a monotonous {\em decrease} of the MI from $0.1$ without noise to almost zero at very large noise levels.

\paragraph*{Motif S3}

In the more strongly connected motif S3 (third row of Fig. \ref{SR-curves}), the state probabilities and the state entropy show a behavior that is qualitatively similar to motif S2. Quantitatively, however, the state entropy without added noise is now with $H(X)\approx 2$ even less optimal than in motif S2: The system is pinned most of the time in only a few dominating states. As the noise level $\sigma$ is gradually increased, the system is freed from these dominating states, and the state entropy $H(X)$ is now growing faster than the stochastic information loss $H(Y|X)$. Since $I(X;Y)=H(X)-H(Y|X)$ in a recurrent network, the MI is now initially {\em increasing} with the noise level, then reaches a maximum at about $\sigma=1$, and eventually falls to zero for even larger noise levels.

\paragraph*{Motif S4}

When investigating the most strongly connected motif S4 (fourth row of Fig. \ref{SR-curves}), we find an even more pronounced maximum of the MI as a function of the noise level.

\newpage

\section*{Discussion}

The phenomenon of stochastic resonance is typically discussed in the context of signal detectors, or sensors, that transmit physical signals from the environment into an information processing system \cite{benzi1981mechanism, wiesenfeld1995stochastic, gammaitoni1998stochastic, moss2004stochastic,krauss2016stochastic,krauss2017adaptive,krauss2018cross}. Many sensors have a detection threshold, that is, a minimum required signal strength below which detection is normally not possible. However, when noise is added to the signal before entering the sensor, even very weak signals can be lifted beyond the threshold with a certain probability. From a theoretical point of view, the ideal way to quantify this effect is by computing the MI between the signal and the sensor output, as this quantity measures the true information transmission across the sensor. If a plot of the MI as a function of the noise level shows a peak, this is considered as the hallmark of stochastic resonance. 

In this work, we have studied the conceptually related, but more complex problem of how noise affects the flux of information in recurrent neural networks (RNN). As our model system we have chosen one of the simplest examples of a RNN, namely motifs of three probabilistic neurons, mutually connected with links of ternary weights. In order to quantify the ongoing information flux in a motif, we used $I(X;Y)$, the MI between successive states. In addition, we considered the state entropy $H(X)$, which measures the average information content circulating in the network, regardless of whether this information is random or temporally correlated. Finally, we considered the stochastic information loss $H(Y|X)$, which is due to the probabilistic nature of the neurons and due to the added noise.

As was to be expected, added noise increases the state entropy $H(X)$ by broadening the distribution of system states within the available state space. At the same time, the noise makes state-to-state transitions more random compared to the undisturbed system, and thus also increases the stochastic information loss $H(Y|X)$. However, since a large information flux requires simultaneously a large state entropy and a small stochastic information loss, the effect of noise depends on the relative rates of increase of $H(X)$ and $H(Y|X)$.

If an undisturbed neural network is already operating in a dynamically rich regime, where all possible system states are visited with approximately the same probability, $H(X)$ cannot be substantially increased by adding noise. On the other hand, the added noise can lead to a rapid increase of the stochastic information loss $H(Y|X)$, in particular if the undisturbed system is behaving in a relatively deterministic way. As a consequence, such neural networks will show a monotonous decrease of $I(X;Y)$ with the noise level. Indeed, we have observed this type of behavior in the weakly connected motif S2.

A qualitatively different behavior is found in neural networks which are originally 'trapped' in a restricted region of state space. In this case, adding just a relatively small amount of noise can quickly 'free' the system from its dynamical trap, leading to a rapid increase of the state entropy $H(X)$. If $H(X)$ is initially growing faster with the noise level than $H(Y|X)$, this will lead to a maximum of $I(X;Y)$. We have found this type of behavior in the more strongly connected motifs S3 and S4.

Thus, our study shows that noise can have either a detrimental or a supporting effect on the information flux in neural networks, and adding noise can be particularly useful in systems that operate within a dynamically sub-optimal regime. If this stochastic resonance like behavior extends to larger neural networks remains to be shown in future studies. Nevertheless, we speculate that the brain could use bursts of noise to free neural network dynamics from being permanently trapped in 'attractor states'.

\newpage

\FloatBarrier

\newpage

\section*{Author contributions}
PK and CM designed the study and developed the theoretical approach. KP performed computer simulations. KP, PK, CM and AS discussed the results. KP and AS prepared the figures. CM and PK wrote the paper. All authors read and approved the final manuscript.

\newpage

\section*{Funding}
The authors are grateful for the donation of two Titan Xp GPUs by the NVIDIA GPU Grant Program.

\newpage

\FloatBarrier

\end{document}